\documentclass[lettersize,journal]{IEEEtran}
\usepackage{amsmath,amssymb,amsfonts}
\usepackage{algorithm}
\usepackage{array}
\usepackage[caption=false,font=normalsize,labelfont=sf,textfont=sf]{subfig}
\usepackage{textcomp}
\usepackage{stfloats}
\usepackage{url}
\usepackage{verbatim}
\usepackage{graphicx}
\usepackage{cite}
\usepackage{comment}%
\usepackage{graphicx}%
\usepackage{multirow}%
\usepackage{amsthm}%
\usepackage{mathrsfs}%
\usepackage{xcolor}%
\usepackage{textcomp}%
\usepackage{manyfoot}%
\usepackage{booktabs}%
\usepackage{algorithm}%
\usepackage{algorithmicx}%
\usepackage{algpseudocode}%
\usepackage{listings}%

\usepackage{hyperref}
\usepackage{booktabs}

\newtheorem{example}{Example}%
\newtheorem{remark}{Remark}%

\begin{document}

\title{A new color image secret sharing protocol}

\author{Jos\'{e} Ignacio Farr\'{a}n
\thanks{J.I. Farr\'{a}n: Departamento de Matematica Aplicada, 
Universidad de Va\-lla\-do\-lid, Spain. 
E-mail: jifarran@uva.es} and 
David Cerezo \thanks{D. Cerezo: Computer Science Enginieer. E-mail: dacema97@gmail.com}
\thanks{Partially supported by MCIN PID2019-105306RB-I00.}
}

\maketitle

\begin{abstract}
Visual cryptography aims to protect images against their possible illegitimate use. 
Thus, one can cipher, hash, or add watermarks for protecting copyright, among others. 
In this paper we provide a new solution to the problem of secret sharing for the case when the secret is an image. 
Our method combines the Shamir scheme for secret sharing using finite fields of characteristic 2 with 
the CBC mode of operation of a secure symmetric cryptographic scheme like AES, so that the security 
relies on that of the mentioned techniques. The resulting shares have the same resolution as that of the original image. 
The idea of the method could be generalized to other multimedia formats like audio or video, adapting the method to 
the corresponding encoded information. 
\end{abstract}



\section{Introduction}\label{sec:Intro}

The purpose of secret sharing schemes in cryptography \cite{Krawczyk} is to divide a secret into pieces among a set of users, 
in such a way that these pieces separately do not have any information about the original secret, and so that 
only authorized sets of participants can retrieve the secret if such users share their pieces, called shares. 
Moreover, a set of users sharing their shares can not obtain the secret if they do not constitute an authorized set. 

The simplest secret sharing schemes are the so-called threshold schemes, where the total number of participants is $n$ and 
there is a minimum number $1\leq t\leq n$, called {\em threshold}\/, so that any set of $t$ or more participants can obtain the secret, 
whereas any set of less than $t$ users can not obtain any information about the secret. The most popular threshold 
schemes are based on the Chinese remainder theorem \cite{Mignotte}, or Lagrange interpolation 
(the Shamir scheme \cite{Shamir}). 
In fact, the Shamir's secret sharing scheme is very simple to apply, and has some extra properties like linearity, 
that allows us to use it in other applications like secure multiparty computation \cite{Cramer}. 

If the secret to share is an image, that is called Visual Secret Sharing Schemes or simply Visual Cryptography, 
there are some papers to build shares of an image with the same purpose. The first paper was by Naor and Shamir \cite{Naor}, 
where their method works only for black an white images, the size of the shares is doubled larger with respect to that of the 
original image, and the shares are grayscale images. Other contributions work for more general images, normally changing also 
the size of the shared images. So the ideas of Naor and Shamir were generalized for grayscale and colour images in \cite{Verheul}, with the same drawbacks as in the previous paper. In \cite{Wu} the secret sharing is done with the aid 
of steganography, namely by using an extra image as covering of the secret image. In the paper \cite{Abdulla} the 
sharing is done by dividing the color channels (Red-Green-Blue) and combining it with a covering extra image. 
Finally, the paper \cite{Mohan} is a survey on visual cryptography with the above techniques. 

On the other hand, the paper \cite{Calkavur} applies the Shamir scheme for color image secret sharing 
by interpreting the pixel as the coefficients of polynomials. The result is that the shares have small size 
than the original image, one has to generate a lot of random constant terms for those polynomials, 
and moreover in order to retrieve the original image one has to solve a lot of linear systems, 
depending on the size of the shared image. Our contribution uses Shamir on the pixels in an alternative way, 
so that the shares have the same size as the original image, we only generate a few random random polynomials, 
and consequently we only have to solve a few linear systems, and expanding the result with the use of a 
symmetric cryptosystem. 

Concerning encryption, in order to cipher an image it is not a good idea to do it with the mode of operation ECB 
(Electronic Code Book), that ciphers the blocks independently, and that allows us to obtain an image similar to the 
original one, just by applying some visual filters \cite{Elashry}. In fact, it is better to use other modes of operation, 
like CBC \cite{Modes}. 

In this paper, we propose a visual secret sharing scheme, on the basis of the Shamir's scheme over a finite field of 
characteristic 2, by combining this scheme with a symmetric encryption scheme like AES \cite{AES} in mode CBC. 
The security of the method is the same as that of the Shamir's scheme, and in addition that of the AES cryptosystem. 
The advantage of our scheme is that it can be applied to any type of image (color or grayscale), and the shares 
have the same size as the original one. Moreover, we claim that the same idea can be used for sharing other types 
of multimedia files, like audio or video. 

The paper is organized as follows: Section 2 recalls some preliminary concepts on digital images and finite fields, 
the threshold scheme of Shamir based on polynomial interpolation, and presents an overview on symmetric cryptography, 
with special emphasis on the Advanced Encryption System (AES) with the Cipher Block Chaining (CBC) mode of operation. 
In the main section 3 we propose a new visual secret sharing scheme for color images, and in section 4 we present 
some examples and experiments, together with some details about our implementation in pure Python~\cite{Python}. 
We end the paper with some conclusions and remarks in section 5.

\section{Preliminares}\label{sec:Pre}

This section is devoted to recall some basics on digital images, finite fields, secret sharing and symmetric cryptography. 
The reader having this background could go directly to section \ref{sec:Scheme}. 

\subsection{Digital images}\label{sec:Images}

Bitmap images consist of a matrix of pixels, where the number of rows is the vertical resolution of the image and the 
number of columns is its horizontal resolution (see~\cite{RGB}). In the RGB system for color images each pixel is encoded by 
three integers, each of them representing the level of (R)ed, (G)reen and (B)lue. 
For example, the Python package scikit-image represents a bitmap color image as a 3D array, 
where the internal arrays are lists of three integers. 

The normal depth of the color channels is 8 bits, so that these integers belong to the interval $[0,255]$\/, 
and hence each pixel is encoded by 24 bits. In the case of grayscale images, each pixel is encoded only 
by an integer representing the corresponding level of gray, normally also in the range $[0,255]$\/, so that 
in this case scikit-image represents a grayscale image just as a 2D array. 

Several formats contain the information about the pixels in raw, that is, without any compression, like BMP, PPM, XPM. 
In practice, there exist many compressed format, with or without loss of information, like JPEG, PNG, GIF or TIFF. 
In this case, an algorithm of decompression reconstructs the original matrix of pixels. 

All the formats have a header containing basic information about the image, like the image type, the resolution, 
the depth of color, etc. The rest of the file contains a stream of bits corresponding to the integers associated 
to the pixels in binary, once the image is read or uncompressed, according to the case. For our purposes, 
we will separate the header on one hand, and on the other hand we will work with the bitstream as a long array of bits. 

\subsection{Finite Fields}\label{sec:FF}

Since the images we will work with are just streams of bits, we are interested in representing the images as 
a sequence of arrays of bits with constant size $m$\/, so that we can identify these bit arrays as elements of 
a finite field of characteristic 2, namely the Galois field with $2^{m}$ elements. 
We recall that such a finite field is constructed from the binary field $\mathbb{F}_{2}=\{0,1\}$ as the field extension 
generated by an irreducible polynomial of degree $m$\/. 

In this way, the elements of $\mathbb{F}_{2^{m}}$ can be regarded as polynomials of degree smaller than $m$ 
with binary coefficients, and thus they can be encoded precisely as arrays of $m$ bits. 
As a consequence, the sum in $\mathbb{F}_{2^{m}}$ can be performed as the bitwise XOR of bit arrays, 
where as the multiplication is usually implemented by a table, containing the correspondence between the 
bit arrays and the powers of a primitive element (see~\cite{McEliece}). 

In case of the irreducible polynomial being a Conway polynomial, the computations are simpler and faster 
(see~\cite{Magma} for further details). However, these Conway polynomials are very hard to compute. 
A list of available Conway polynomials can be seen in~\cite{Conway}. For the sake of efficiency, 
the possible lengths of bit arrays are restricted to correspond to the degree of a Conway polynomial in characteristic 2. 
In our Python implementation we have chosen $m=64$, so that the chosen Conway polynomial is 
\begin{multline*}
P(X) = X^{64} + X^{33} + X^{30} + X^{26} + X^{25} + X^{24} + X^{23} + \\
X^{22} + X^{21} + X^{20} + X^{18} + X^{13} + X^{12} + X^{11} + \\
+ X^{10} + X^7 + X^5 + X^4 + X^2 + X + 1 \\
\end{multline*}

\vspace{-.5cm}

\noindent This field is efficiently implemented in Python with the aid of the package {\tt galois}. 
The use of finite fields is required because of the fact that, in the next paragraph, we need to perform linear algebra 
in order to solve systems of linear equations, so that we cannot just use integer or modular arithmetic. 

\subsection{Shamir secret sharing}\label{sec:SSS}

When a piece of information is too valuable to be accessible to an only person, there exist protocols to divide this 
secret information among several parts, so that only authorized sets of participants can get the secret 
if they share their parts of the information. These protocols are called secret sharing schemes. 
The simplest secret sharing protocols are called $(t,n)$-threshold schemes, where the secret information 
can be retrieved if at least $t$ out of $n$ participants share their partial informations (called shares), 
so that any set with less than $t$ participant cannot get any information about the original secret. 

The Shamir secret sharing scheme \cite{Shamir} is the one that is used the most in many situations, 
because its linearity, meaning that a linear combination of some secrets can be obtained 
as the corresponding linear combination of the shares. The Shamir scheme works as follows: 

\begin{itemize}

\item The secret $S$ is a number, and it is hidden into a polynomial $f(x)$ of degree $t-1$. 
For example, $S$ can be the constant term, that is the polynomial evaluated at the origin, 
and the other coefficients of $f$ are generated at random. 

\item Each participant has a distinct identification number $I$, and then the share given to this 
participant is precisely $f(I)$. 

\item Since the polynomial $f$ can be determined by the values at $t$ different points, if $t$ participants 
share their couples $(I,f(I))$ they can determine $f$, and hence $S=f(0)$, by simple Lagrange interpolation 
over a (finite) field. Notice that, since the identification numbers are different, the matrix of the 
corresponding linear system is of Vandermonde type, so that the solution always exists and it is unique. 

\item Note also that less than $t$ shares lead to an underdetermined system, so that there are at least 
as many solutions as elements in the corresponding base field, and hence a brute-force attack is not feasible 
as long as the size $q$ of the underlying finite field $\mathbb{F}_{q}$ is large enough. 

\end{itemize}

In our case, we apply the above scheme over the finite field $\mathbb{F}_{2^{64}}$\/, 
so that the identification numbers, as well as the random coefficients, are all elements of this field. 

\medskip 

\begin{example}

We share the secret number $S=1234$ with the Shamir $(3,6)$-threshold scheme, over the field of real numbers. 
We generate at random the polynomial 
\[
P(x)=1234+166x+94x^{2}
\]
and distribute the shares 
\[
(1, 1494),(3, 2578),(4, 3402),(6, 5614),(8, 8578),(11, 14434)
\]
Now, if the users 2, 5 and 6 join their shares, they obtain the linear system 
\[
\left\{
\begin{array}{rcrcrcl}
s&+&3a_{1}&+&9a_{2}&=&2578\\
s&+&8a_{1}&+&64a_{2}&=&8578\\
s&+&11a_{1}&+&121a_{2}&=&14434
\end{array}
\right.
\]
whose unique solution determines $P(x)$, obtaining then the secret $S=P(0)=1234$. 


\end{example}

\subsection{AES and CBC}\label{sec:AES}

The last ingredient of our algorithm is the use of a secure enough symmetric cryptosystem 
like the Advanced Encryption Algorithm (AES). The standard specification of AES is given 
by the Federal Information Processing Standards Publication 197 \cite{AES}. 
This is a block cipher, operating on blocks of bits with size 128, and using keys of bit size 
128, 192 or 256 (AES128, AES192 and AES256, respectively). 
There is no known efficient attack to AES, and the security increases as the key size is larger. 
Moreover, the algorithm is efficient in time and space. 
The details of the algorithm is referred to the cited FIPS 197 publication \cite{AES}. 

The idea of our algorithm is to use the first bits of the image as a key for the AES encryption, 
to apply the Shamir secret sharing protocol to share the key, and encrypt the rest of the image with this key. 
The problem is that, when ciphering an image with the ECB mode of operation, 
the aspect of the encrypted image shows patterns that can be seen with the aid of some visual filters 
(see~\cite{Elashry} for further details). Thus, it is better to use alternative modes of operations \cite{Modes}, 
like CBC, CFB or OFB. In all these three cases we need an initialization vector, that can be computed from 
the identification number $i$ used in the Shamir scheme. This makes the shared images completely different 
for each of the participants. 

In our Python implementation we have chosen the CBC mode, with the AES128 version, but 
these are parameters that can be changed in order to optimize the security of the algorithm. 
If we want to use the AES192 or the AES256 version, then we need to use the corresponding 
number of bits of the original image as the key of the algorithm. 

In order to increase the security, instead of choosing the first bits of the image as key for the AES cipher, 
we could used the identification number $i$ (or a hash of it) to set the position of the consecutive bits that are used 
for the cipher key. 

\section{The proposed visual secret sharing scheme}\label{sec:Scheme}

First we describe the process of generating the shares of a given image, for a fixed finite field $\mathbb{F}_{2^{m}}$\/. 
Since we are going to use AES and the corresponding key is going to be shared, the optimal $m$ would be 
either 128 or 256. Since a Conway polynomial for these finite fields is not available, an alternative is to use 
$m= 64$, as we saw in section \ref{sec:FF}. Denote by $M$ the number of bits of the AES version we choose, 
that is $M\in\{128,192,256\}$\/, and assume from now on that $m=64$. In case we can handle the finite fields 
with $m=128$ or $m=256$, the algorithms can be easily adapted to the cases $M=128,256$. 

In all the cases, we are using the CBC or the OFB modes of operation with AES (see~\cite{Modes}). 
Then we need an initialization vector IV of 128 bits for the encryption, that for each participant will be 
the first 128 bits of a secure hash like SHA \cite{SHA} of the corresponding identification number $I$, 
generated in the Shamir secret sharing. In the case that we use the finite field with $m=128$, 
this hashing is not necessary. 

Assume that the information about the pixels is stored in an array $B$ of bits, 
and that the number $N$ of bits of $B$ is a multiple of 128, since AES works on blocks of bits of length 128. 
If this is not the case, the last remaining bits of $B$ may not be ciphered, 
since the information they contain is not relevant compared to the whole image. 
Nevertheless, since normally each pixel is encoded by 24 bits, the above condition is easily satisfied as long as 
the total number of pixels is a multiple of 16; this is the case when both the horizontal and vertical resolution 
are multiples of 4. Otherwise we can fill the original image with 
random pixels at the margins in order to get this condition, and add this information to the file header. 
Other alternative is to use the CFB mode \cite{Modes}, adapting the size of the blocks in $B$. 

On the other hand, we need to save apart the essential information contained in the header of the image, 
namely the horizontal and vertical resolution, and the depth of color channels 
(by default 255, meaning that each pixel is represented by 24 bits). 
This will be necessary to retrieve the original image from the shares. 

Thus, assume in the sequel that there are $n$ participants, and that the threshold is $t$. 
Denote $q=2^{64}$, and $M=64\ell$. Note that a block of 64 bits can be interpreted as 
an element of $I\in\mathbb{F}_{q}$ when we are applying the Shamir secret sharing method. 
The procedure to generate the shares is described in Algorithm~\ref{shadows}. 

\begin{algorithm}
\caption{Procedure to generate the shares of a color image.}
\label{shadows}

$\,$ 

{\bf Input:} The bit array $B$ of $N$ bits, the length $M$, and the parameters $n$ and $t$. \par
\medskip 
\begin{itemize}
\item For each participant $i$\/: 
\begin{enumerate}
\item Generate at random an identifier $I\in\mathbb{F}_{q}$\/. 
\item Compute $\mathrm{IV}^{i}$ from the hash SHA of $I$. 
\end{enumerate}
\item Get $K$ the first $M$ bits of $B$. \\
\item Divide $K$ into $\ell$ pieces of 64 bits $M_{j}$\/. \\
\item For each $M_{j}$\/: generate $n$ shares $S^{i}_{j}$ with the Shamir secret sharing, 
i.e. $S^{i}_{j}=f_{j}(I)$ for a certain random polynomial $f_{j}$ of degree $t-1$ with $f_{j}(0)=M_{j}$\/. \\
\item For each participant generate the image-share $C^{i}$ as follows: 
\begin{enumerate}
\item The first $M$ bits are the concatenation of the shares $S^{i}_{j}$\/. 
\item The remaining bits are the result of applying AES in CBC/OFB mode to the last $N-M$ bits of $B$, 
with the key $K$ and initialization vector $\mathrm{IV}^{i}$\/. 
\end{enumerate}
{\bf Output:} The $n$ shares $C^{i}$ together with the identifiers $I$. 
\end{itemize}

\end{algorithm}

\medskip 

\begin{remark}\label{steg}

In order to increase the security, instead of taking the first $M$ bits of $B$ as the key $K$ for the AES encryption, 
we could take $M$ consecutive bits from a random position of $B$. 
This position should be known for all he participants in order to reconstruct the original image. 

In this case we remove these bits from $B$, cipher the remaining bits as in Algorithm~\ref{shadows}, 
and place the shares $S^{i}_{j}$ concatenated at the beginning of $C^{i}$\/. 
The random position of the key in $B$ must be kept in mind in order to insert the bits of $K$ at the right position 
in the original image. 

In fact, the position where the selected bits start can be dependent on the identifier $I$, in terms of a hash for instance, 
so that this position is different for each user in the scheme. 

\end{remark}

\medskip 

Now we are going to describe the process to retrieve the original image $B$ from the shares $C^{i}$ 
of at least $t$ participants. Each of the $t$ participants must share with the others the identifier $I$ 
and the first $M$ bits of $C^{i}$\/, that is, the concatenation of $S^{i}_{j}=f_{j}(I)$\/. 
The procedure of reconstructing the original image is described in Algorithm~\ref{retrieve}. 

\begin{algorithm}
\caption{Procedure to reconstruct the original color image.}
\label{retrieve}

$\,$ 

{\bf Input:} The data $I$ and $C^{i}$ of $t$ participants. \par
\medskip 
\begin{itemize}
\item Each user shares $I$ and the first $M$ bits of $C^{i}$\/. 
\item Slice the above bits to obtain the shares $S^{i}_{j}=f_{j}(I)$\/.  
\item Retrieve $K$ by $\ell$ interpolation procedures, as in the Shamir method. 
\item Each of the $t$ participants reconstruct the original image $B$ as follows: 
\begin{enumerate}
\item The first $M$ bits of $B$ are those of $K$. 
\item Compute $\mathrm{IV}^{i}$ from the hash SHA of $I$. 
\item The remaining bits are the result of deciphering with AES in CBC/OFB mode the last $N-M$ bits of $C^{i}$, 
with the key $K$ and initialization vector $\mathrm{IV}^{i}$\/. 
\end{enumerate}
{\bf Output:} The original color image $B$\/. 
\end{itemize}

\end{algorithm}

\bigskip 


\begin{remark}

Following the idea of remark~\ref{steg}, if the selected bits for the encryption key are not at the beginning of the original image, 
we have to insert these bits in the correct position, once we have deciphered the tail of the bitarray. 

\end{remark}

\begin{remark}

Note that the generated shared images have the following properties: 

\begin{enumerate}

\item The shares have the same resolution as the original images, and they are all color images. 

\item All the shares look random because of the CBC/OFB mode operation of AES, so that 
no information can be extracted from them without the key $K$ used for the encryption. 

\item All the shares are completely different, since in the encryption distinct initialization vectors 
are used by each participant. 

\end{enumerate}

Finally, we note that the security of these procedures relies of the Shamir secret sharing protocol, 
together with that of the AES cipher and the corresponding mode of operation. 

\end{remark}

\section{Examples and experiments}\label{sec:Ex}

We have a Python implementation of the algorithms available in \cite{git}
\begin{quote}
\href{https://github.com/jifarran/color_visual_crypto}{https://github.com/jifarran/color\_visual\_crypto}
\end{quote}
The required packages that we have used for the implementation are, among others 
{\tt skimage}, {\tt bitarray}, {\tt galois}, {\tt numpy} and {\tt Crypto}. 

\begin{example}

As an example, you can see a test image in figure~\ref{bridge}, 
and the shared images by a $(3,2)$-threshold scheme in table~\ref{tab-shares}. 
On the other hand, you can see in figure~\ref{pont} the reconstruction of the original imagen from the first two shares. 

\begin{figure}[h]
\centering
\includegraphics[width=0.3\textwidth]{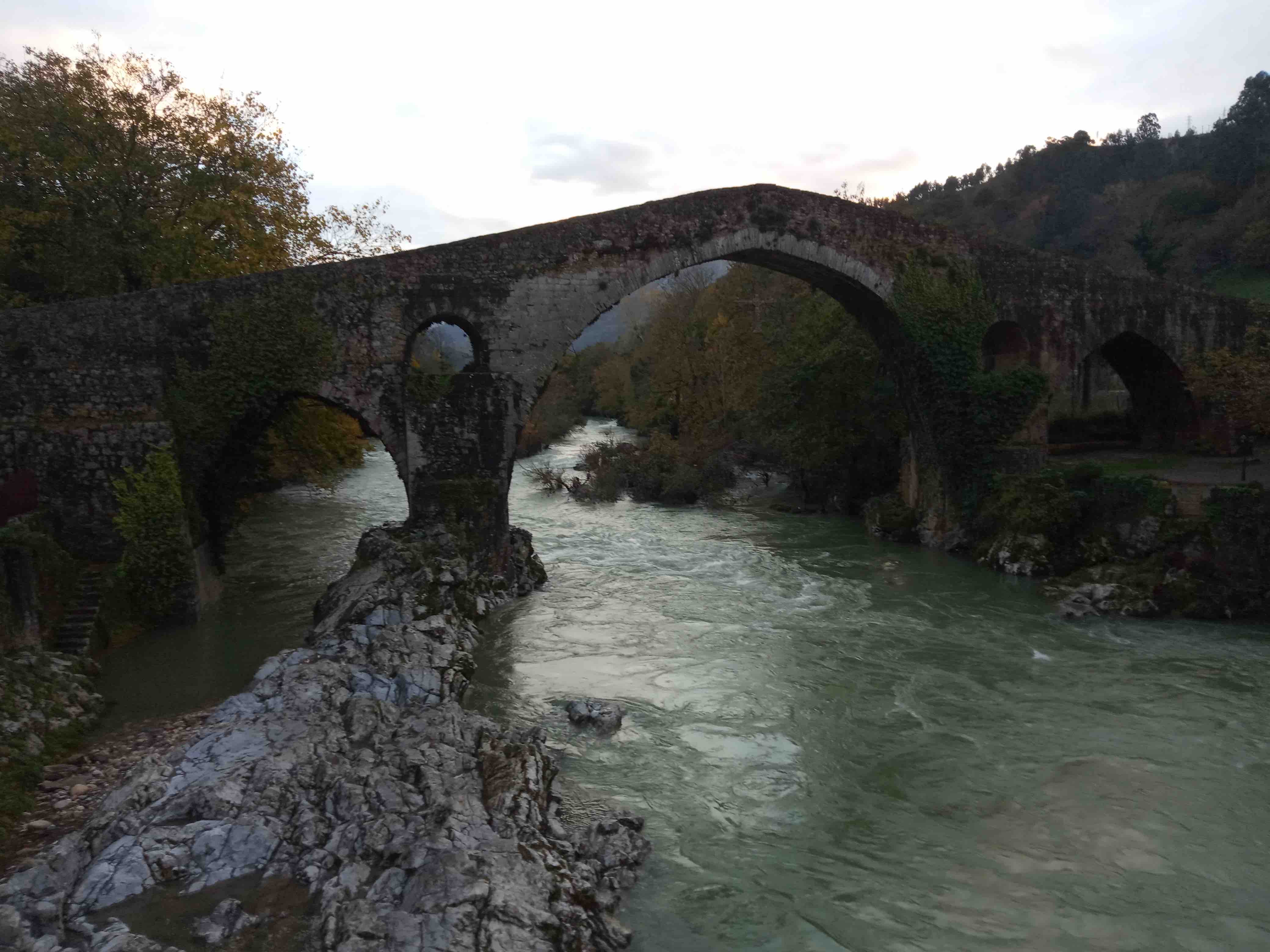}
\caption{Original image to be shared.}\label{bridge}
\end{figure}

\begin{table}[h]
\centering
\begin{tabular}{|c|}
\toprule
\includegraphics[width=0.3\textwidth]{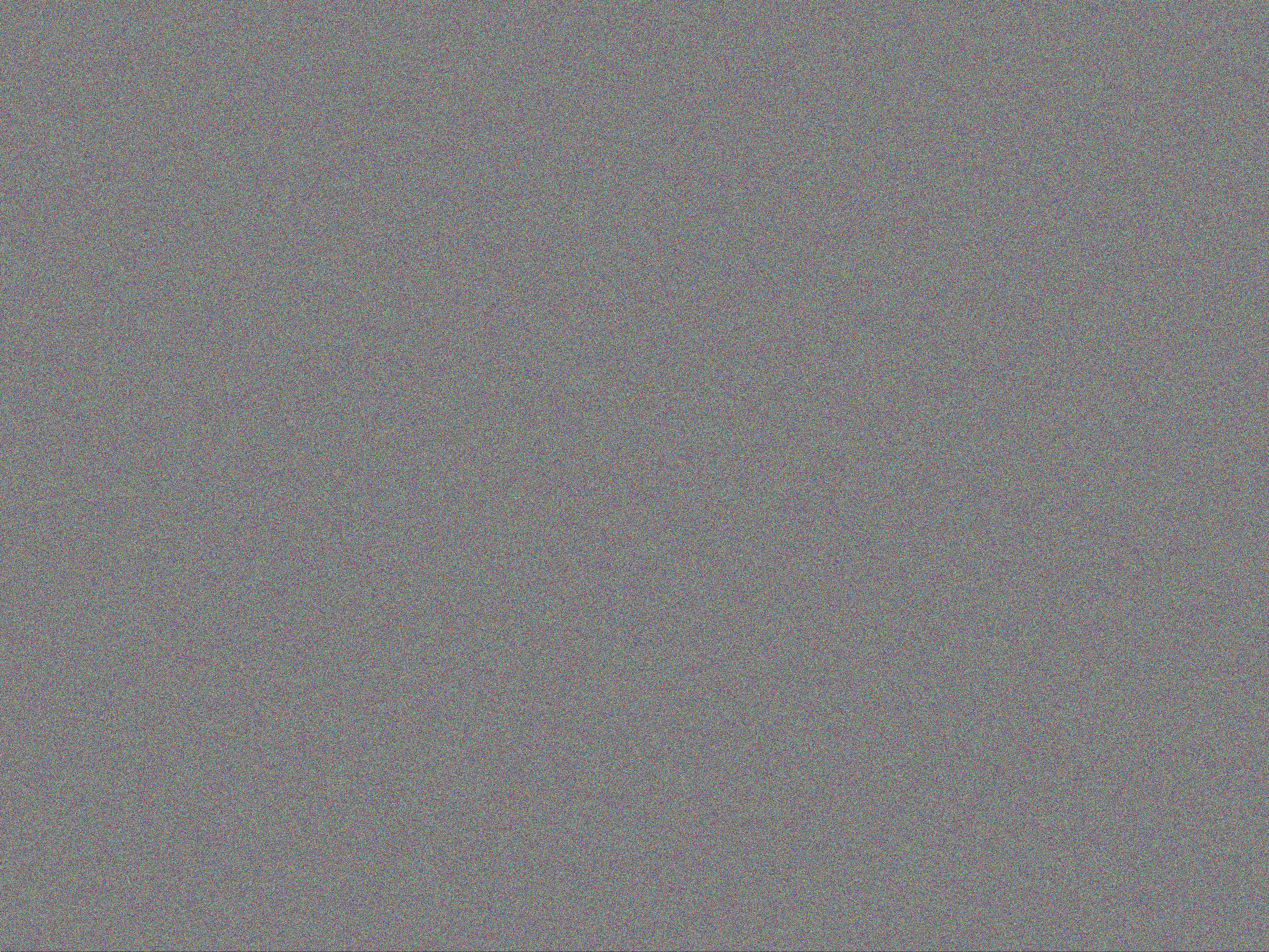} \\
share 1 \\
\midrule
\includegraphics[width=0.3\textwidth]{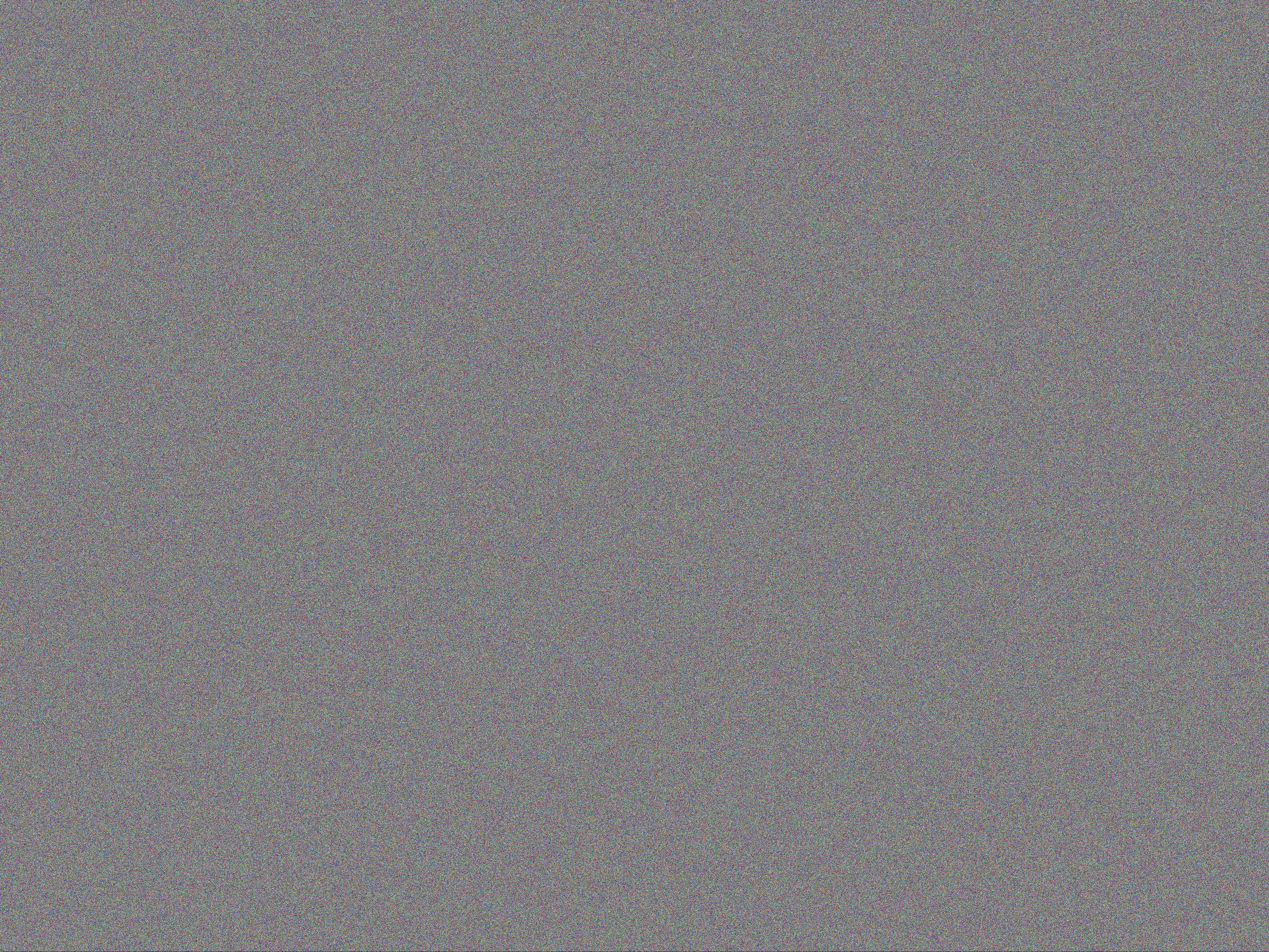} \\ 
share 2 \\
\midrule
\includegraphics[width=0.3\textwidth]{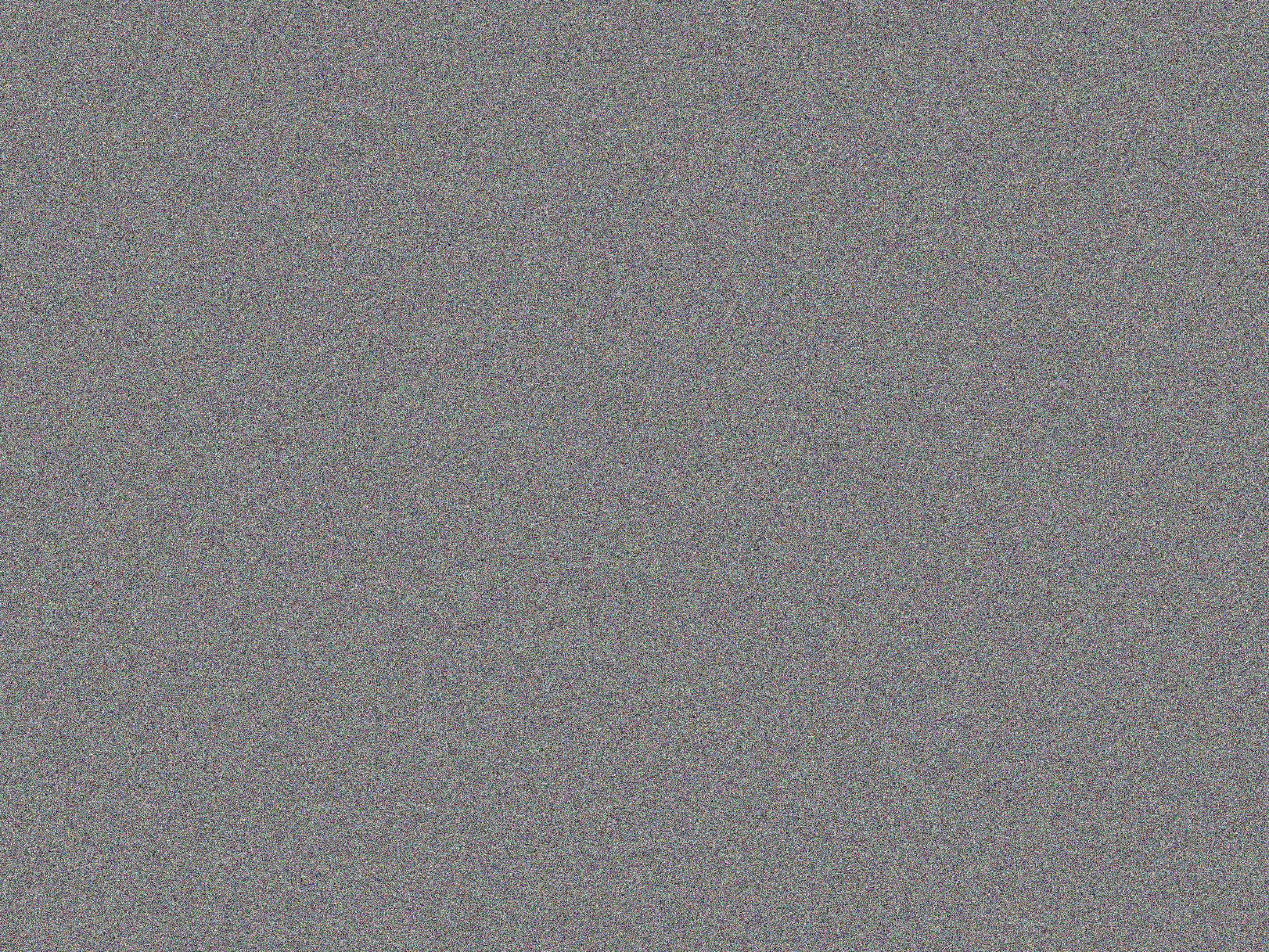} \\
share 3 \\
\bottomrule
\end{tabular}
\caption{Shares of image in figure \ref{bridge}.}\label{tab-shares}
\end{table}

\begin{figure}[h]
\centering
\includegraphics[width=0.3\textwidth]{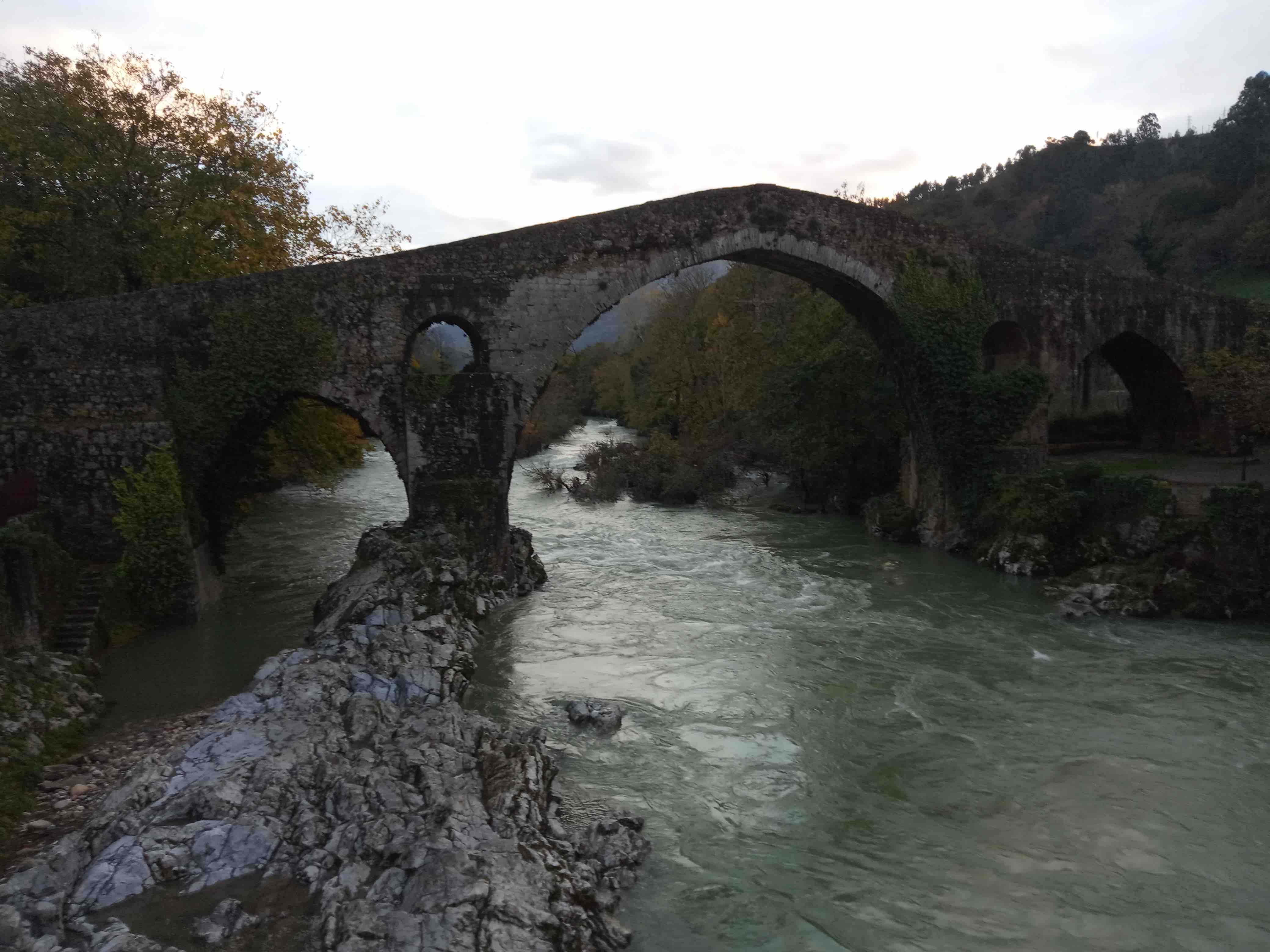}
\caption{Reconstructed image from two shares.}\label{pont}
\end{figure}

\end{example}


We have done several tests with images of different size, 
on a MAC with processor 2,3 GHz Intel dual Core i5 and 8GB of RAM. 
In tables \ref{timingsKS} and \ref{timingsKR} the sizes of the images are expressed in Megabytes on the left column. 
The execution times to compute the image shares 
are shown in table~\ref{timingsKS}. 
In our implementation, $M=128$, $q=2^{64}$, we use AES128 in CBC mode, 
and we do not use the hashing SHA, but for the sake of simplicity 
we concatenate twice the identifier $I$ in algorithm~\ref{shadows} instead. 

\begin{table}[h]
\centering
\begin{tabular}{|r||r|r|r|r|}
\toprule
(t,n) & (2,3) & (3,5) & (5,5) & (6,10) \\
\hline\hline 
4.9MB & 51.07s. & 84.37s. & 85.31s. & 166.03s. \\
\hline 
9.4MB & 101.02s. & 160.87s. & 164.38s. & 308.05s. \\
\hline 
24.9MB & 277.38s. & 419.96s. & 422.41s. & 840.93s. \\
\hline 
49.2MB & 564.61s. & 893.06s. & 885.27s. & 1813.78s. \\
\bottomrule
\end{tabular}
\caption{Computation times to obtain the shares of the test images.}\label{timingsKS}
\end{table}

Note that the computation time to obtain the shares $S_{j}^{i}$ only depends on the parameters 
$n$, $t$ and $M$ in the algorithm \ref{shadows}, so that it will be irrelevant compared to the 
encryption time, that depends on the size of the image. 
Thus, from the results of this table, we can conclude that the time to compute the image shares 
seems to be linear with respect to the size of the uncompressed data. 
In fact, on our computer turns out to be approximately around 3.5 seconds per MB of data 
(note that in each case we have to obtain $n$ images). 
The total times includes the time to open the file, do some internal conversions, and apply the AES in CBC mode. 

We show now the computation times for the reconstruction of the original image in table~\ref{timingsKR}. 
Note that, in this case, we only have to reconstruct one image, and not $n$ images as in the sharing algorithm. 

\begin{table}[h]
\centering
\begin{tabular}{|r||r|r|r|r|}
\toprule
(t,n) & (2,3) & (3,5) & (5,5) & (6,10) \\
\hline\hline 
4.9MB & 24.80s. & 33.32s. & 53.60s. & 64.25s. \\
\hline 
9.4MB & 43.05s. & 62.54s. & 105.71s. & 125.17s. \\
\hline 
24.9MB & 119.90s. & 174.23s. & 289.60s. & 346.43s. \\
\hline 
49.2MB & 263.79s. & 371.40s. & 614.21s. & 800.00s. \\
\bottomrule
\end{tabular}
\caption{Computation times to obtain the originals for the test images.}\label{timingsKR}
\end{table}

Note that, in this case, by comparing the cases $(t,n)=(3,5)$ and $(t,n)=(5,5)$, 
the parameter $t$ has an influence in the time of computation, 
since we have to solve a linear system of size $t$ in the finite field $\mathbb{F}_{q}$, 
and this has a complexity $\mathcal{O}(t^{3})$ with Gaussian elimination. 
In fact, most of the time in algorithm \ref{retrieve} is devoted to the interpolation task 
in Shamir secret sharing procedure. 

In this case, the time of computation does not seem to be linear with respect to the size 
of the data, but one easily checks that the computation time per MB of data do not differ very much, 
for $t$ and $n$ fixed. Anyway, we have experimented with large images, but for small images 
the times of computation are quite reasonable, although we could optimize a lot the implementation to speed-up the algorithms. 


\begin{remark}[File formats]

We finally notice that, in our implementation, two formats of images are admitted: PPM and PNG. 
In the second case there is a lossless compression, while in the first case there is no compression at all. 
Lossy compression make no sense in this context, since this type of compression may downgrade the 
quality of the image. 

All the above computation times are calculated with respect to the size of the uncompressed image, 
although in the implementation the shares and the reconstructed images are generated in PNG format, 
in order to save space. The {\tt skimage} Python package automatically compresses and uncompresses 
the corresponding images, without loss of information. 

\end{remark}

\section{Conclusions}\label{sec:Fin}

A new and efficient method to perform secret sharing with color image has been presented, on the basis of 
the Shamir threshold secret sharing scheme over binary finite fields, combined with a secure symmetric cipher like AES 
in CBC/OFB mode of operation. Thus, the security of the method is based on that of the above techniques, 
which have been proven secure so far. 

The advantages of our method, compared to the existing ones, are the following: 

\begin{enumerate}[(1)]
\item The shares are images of the same type (color images). 
\item The shares have all the same resolution. 
\item The shares look quite random, showing no visual pattern. 
\item The shares are completely different each other. 
\end{enumerate}

Concerning future work, the authors propose to apply the same idea to do secret sharing with other types of 
multimedia or binary formats (audio, video, etc.). In the case of video, or other kinds of lossy compressed data, 
one should study carefully the impact of the loss of information in the secret sharing process, 
but if the compression is reliable enough, our method may work properly.

\newpage

\begin{IEEEbiographynophoto}{J. I. Farr\'{a}n}

He received his PhD in Mathematics from the Unviersity of Valladolid, Spain, in 1997. 
Since 2001 his is an associate professor at the Department of Applied Mathematics, in the University of Valladolid, 
being a lecturer in the Computer Science School of Segovia, and a member of the IMUVa 
(Institute for Mathematic Research of the University of Valladolid). 
His research interests are in algebraic geometry, computer algebra, numerical semigroups, information and coding theory, 
cryptography, and analysis of meteorological data. He is a contributor to the computer algebra system {\sc Singular}, 
from the Technische Universit\"{a}t Kaiserslautern. 

\end{IEEEbiographynophoto}


\begin{IEEEbiographynophoto}{D. Cerezo}
He received his degree in Computer Science Engineering from the University of Valladolid, Spain, in 2022. 
He is an expert in cybersecurity, network security and ethical hacking. 
\end{IEEEbiographynophoto}

\vfill

\end{document}